\documentclass{article}

\PassOptionsToPackage{numbers,sort&compress}{natbib}
\usepackage[sglblindworkshop, final]{ai4nextg_neurips_2025}

\usepackage[utf8]{inputenc} 
\usepackage[T1]{fontenc}    
\usepackage{hyperref}       
\usepackage{url}            
\usepackage{booktabs}       
\usepackage{amsfonts}       
\usepackage{nicefrac}       
\usepackage{microtype}      
\usepackage{xcolor}         
\usepackage{soul}           

\usepackage{times}
\usepackage{amsmath,amsfonts,amssymb}
\usepackage{booktabs}
\usepackage{algorithm}
\usepackage{algorithmic}
\usepackage{array}
\usepackage[caption=false,font=normalsize,labelfont=sf,textfont=sf]{subfig}
\usepackage[justification=raggedright]{caption}
\usepackage{textcomp}
\usepackage{stfloats}
\usepackage{verbatim}
\usepackage{graphicx}
\usepackage{subfig}
\usepackage{wrapfig}
\usepackage{tabularx}
\usepackage{enumitem,kantlipsum}
\usepackage{hyperref}
\usepackage{cleveref}
\crefname{figure}{Fig.}{Figs.}
\Crefname{figure}{Fig.}{Figs.}
\crefname{equation}{Eq.}{Eqs.}
\Crefname{equation}{Eq.}{Eqs.}
\usepackage{booktabs,tabularx,makecell}

\renewcommand{\arraystretch}{1.15}
\newif\ifcompact \compacttrue
\ifcompact
  \renewcommand{\arraystretch}{0.95}    
\fi

\ifcompact
\fi

\ifcompact
  \AtBeginDocument{%
    \setlength{\abovedisplayskip}{6pt plus 2pt minus 2pt}%
    \setlength{\belowdisplayskip}{6pt plus 2pt minus 2pt}%
    \setlength{\abovedisplayshortskip}{3pt plus 2pt minus 1pt}%
    \setlength{\belowdisplayshortskip}{3pt plus 2pt minus 1pt}%
    \jot=2pt 
  }

\title{Cross-Layer Design for Near-Field mmWave Beam Management and Scheduling under Delay-Sensitive Traffic}

\author{%
  Zijun Wang\\
  Department of Electrical Engineering\\
  University at Buffalo\\
  Buffalo, NY 14226 USA \\
  \texttt{zwang267@buffalo.edu} \\
  \And
  Anjali Omer \\
  Department of Electrical Engineering\\
  University at Buffalo\\
  Buffalo, NY 14226 USA \\
  \texttt{anjaliom@buffalo.edu} \\
  \AND
  Jacob Chakareski \\
  Ying Wu College of Computing\\
  New Jersey Institute for Technology \\
   Newark, NJ 07103 USA \\
  \texttt{jacobcha@njit.edu} \\
  \And
  Nicholas Mastronarde \\
  Department of Electrical Engineering\\
  University at Buffalo\\
  Buffalo, NY 14226 USA \\
  \texttt{nmastron@buffalo.edu} \\
  \And
  Rui Zhang \\
  Department of Electrical Engineering\\
  University at Buffalo\\
  Buffalo, NY 14226 USA \\
  \texttt{rzhang45@buffalo.edu} \\
}

\begin{document}

\maketitle
\begin{abstract}\label{sec:intro}
Next-generation wireless networks will rely on mmWave/sub-THz spectrum and extremely large antenna arrays (ELAAs). This will push their operation into the near-field where far-field beam management degrades and beam training becomes more costly and must be done more frequently. Because ELAA training and data transmission consume energy and training trades off with service time, we pose a cross-layer control problem that couples PHY-layer beam management with MAC-layer service under delay-sensitive traffic. The controller decides when to retrain and how aggressively to train (pilot count and sparsity) while allocating transmit power, explicitly balancing pilot overhead, data-phase rate, and energy to reduce the queueing delay of MAC-layer frames/packets to be transmitted. We model the problem as a partially observable Markov decision process and solve it with deep reinforcement learning. In simulations with a realistic near-field channel and varying mobility and traffic load, the learned policy outperforms strong 5G-NR–style baselines at a comparable energy: it achieves 85.5\% higher throughput than DFT sweeping and reduces the overflow rate by 78\%. These results indicate a practical path to overhead-aware, traffic-adaptive near-field beam management with implications for emerging low-latency high-rate next-generation applications such as digital twin, spatial computing, and immersive communication.

\end{abstract}

\section{Introduction}
Millimeter-wave (mmWave) and sub-THz bands offer abundant spectrum for high-speed links, while extremely large antenna arrays (ELAAs) are employed to overcome the associated high path loss \cite{6G-tutorial,6G-tutorial2,terahertz-tutorial,mimo-tutorial}. However, the use of ELAAs means that some uses, traditionally assumed to operate in the far-field regime, now fall within the near-field region \cite{Tutorial_with_beam_pattern_and_resolution, General_tutorial, fra_and_ray_2, baduge2025frequencyrange3isac}.
Conventional far-field design models electromagnetic waves as planar \cite{far-field_assumption}. Under this assumption, array responses vary primarily by angle-dependent phase shifts, enabling the channel to be compactly represented using a discrete Fourier transform (DFT) basis, with sparsity roughly corresponding to the number of dominant paths \cite{far-field-assumption2}. In the near field, wavefronts are spherical, and the steering vector becomes a function of both angle and range. As a result, the channel's DFT-domain representation is no longer sparse in a way that simply reflects the number of paths; instead, its sparsity varies with user geometry and mobility \cite{wang2025lowcomplexitynearfieldbeamtraining, channel_model2}.
This undermines the use of far-field codebooks and increases the risk of misalignment. Near-field codebooks have been explored \cite{Cui_and_Dai_channel_model}, yet larger dictionaries and codeword correlation increase training time and energy overhead that directly competes with data transmission. A pragmatic alternative is to combine 5G-new radio (NR)-style sweeping \cite{qin2023reviewcodebookscsifeedback} with near-field codebooks to avoid explicit sparsity selection \cite{wang2025sparsityawarenearfieldbeamtraining}; still, the sweeping/reporting overhead can exceed that of standard NR due to the expanded search space.

Critically, beam training is deeply intertwined with MAC-layer service scheduling \cite{leiJointBeamTraining2022}. Each pilot transmission consumes airtime and energy that could otherwise be used to serve queued traffic. Under bursty arrivals, finite buffers, and mobility-induced dynamics, a PHY-only design that optimizes instantaneous link metrics can still hurt end-to-end latency. This motivates a cross-layer treatment that co-optimizes training timing and intensity with power allocation and queue-aware scheduling \cite{MastronardeSC:21,SharmaMC:18a}.

Prior work has partially addressed this coupling in directional and mmWave systems. Shokri-Ghadikolaei et al.\ formalize the alignment–throughput trade-off and propose joint beamwidth/scheduling strategies \cite{shokrighadikolaeiBeamsearchingTransmissionScheduling2015}, while subsequent work coordinates beam schedules with mobility and sleep/wake cycles to target energy, delay, or throughput \cite{huangBeamAwareCrossLayerDRX2020, panBeamAwareScheduling5G2023, liLocationawareDynamicBeam2018}. 
Lei et al.\ show that adaptive retraining and power control can significantly improve delay/energy compared with fixed policies \cite{leiJointBeamTraining2022}. However, these studies primarily assume far-field propagation and do not tackle near-field-specific issues: angle–range coupling, expanded codebooks, and variable sparsity that governs training intensity.

This paper formulates near-field beam management as a cross-layer control problem for delay-sensitive traffic with minimal energy consumption. We propose a queue-aware policy that jointly decides when to retrain and how aggressively to train (pilot budget and sparsity level), together with data-phase power allocation. Our implementation incorporates compressive-sensing-based training and a deep reinforcement learning (DRL)-based controller that observes queue states and recent training history to balance pilot overhead, service rate, and energy. In simulations with near-field channels over a range of mobility and load models, the learned policy reduces queueing delay and overflow at a comparable energy to strong baselines. The proposed approach narrows the gap to full-channel state information (CSI) performance while offering an overhead-aware and traffic-adaptive solution. Our advances can have implications for emerging low-latency high-rate next-generation applications such as digital twin, spatial computing, and immersive communication that increasingly integrate mmWave capabilities \cite{GuptaCP:20,BadnavaCH:24a,ChakareskiKRB:21,ChakareskiK:23,SrinivasanSAC:24}.

\section{System Model}\label{sec:system model}
In this section, we discuss the channel model, beam training method and data queuing model, which are essential for understanding our cross-layer decision model presented in Section~\ref{sec:ppo}.
\subsection{Channel Model}\label{subsec:channel model}
\begin{figure}[h]
  \centering
 \includegraphics[width=2.5in]{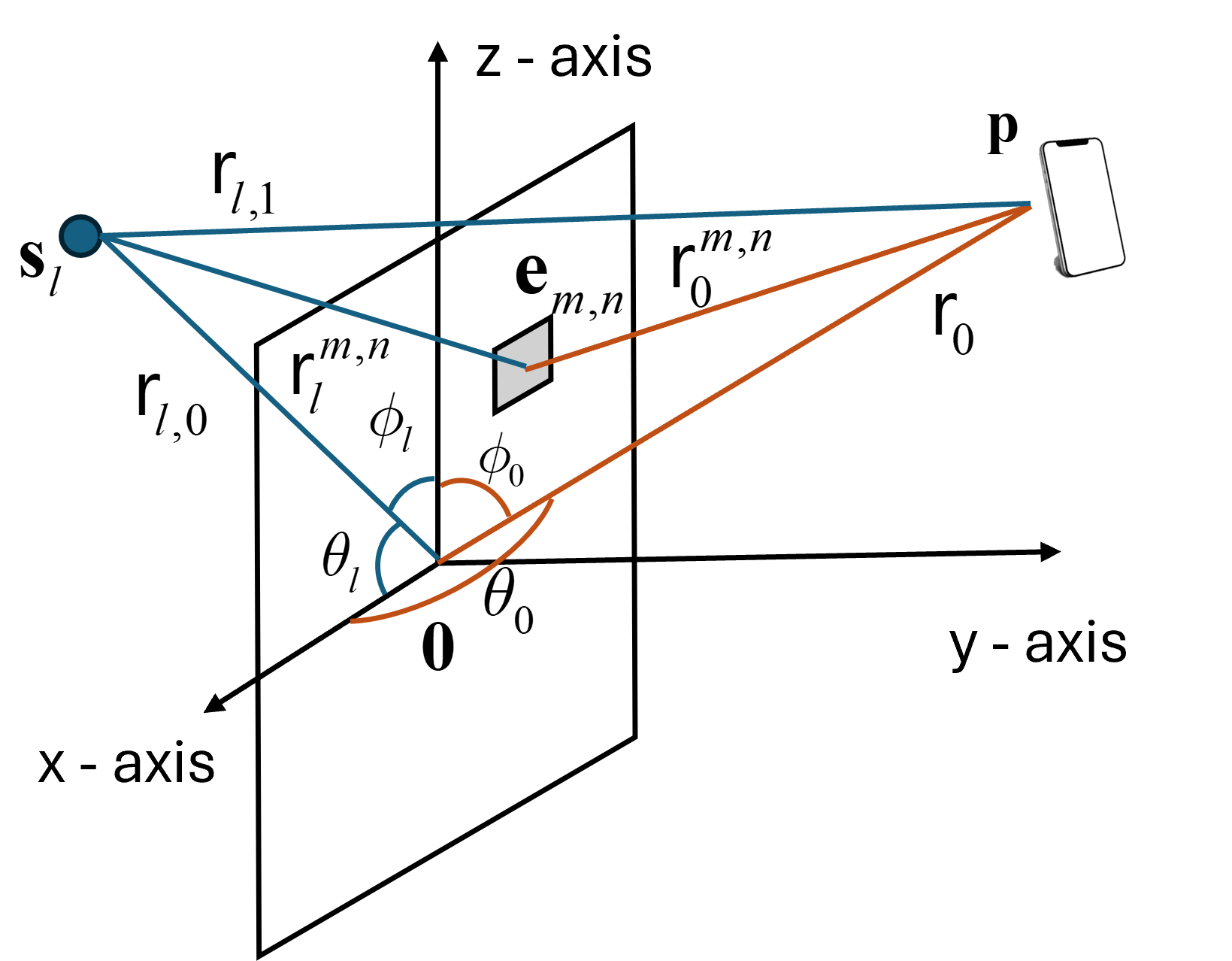}
  \caption{Near-field communication system with UPA.}
  \label{fig:upa}
\end{figure}

We consider a narrow band multiple-input single-output (MISO)  mmWave communication system as shown in \Cref{fig:upa}, where the gNB is equipped with a uniform planar array (UPA). The UPA is placed on the $x-z$ plane and the center of the UPA is at $\mathbf{0}=(0,0,0)$. The number of antenna elements of the UPA is $M = M_z\times M_x$, where $M_z$ and $M_x$ are the number of elements in the $z$ and $x$ directions, respectively. For a UPA, the near-field region lies between the Fresnel distance $R_{Fre}=\frac{1}{2} \sqrt{\frac{D^{3}}{\lambda}}$ and the Rayleigh distance $R_{Ray}=\frac{2D^2}{\lambda}$, where $\lambda$ is the wavelength at the central frequency, $D=\sqrt{(M_xd)^2+(M_zd)^2}$ is the aperture of the UPA \cite{Fraunhofer_and_Fresnel_Distances}, and $d=\frac{\lambda}{2}$ is the spacing of the antenna elements. The far field lies past $R_{Ray}$. In this paper, we focus on cross-layer beam management and data transmission of users in the near field region between $R_{Fre}$ and $R_{Ray}$.

We consider a multi-path ray-tracing channel model as shown in \Cref{fig:upa}. Suppose that the user is positioned at $\mathbf{p}$, and the distance from the center of the antenna to the user is $r_0=\|\mathbf p\|$. Then the distance $r_0^{m,n}$ between the $(m,n)$-th antenna element with coordinate $\mathbf{e}_{m,n}$ and the user is 
\begin{align}
r_{0}^{\,m,n} &=\big\|\mathbf{p}-\mathbf{e}_{m,n}\big\|
= \sqrt{\,r_0^{2}+\delta_{n}^{2}d^{2}+\delta_{m}^{2}d^{2}
         -2r_0\delta_{n}\cos\theta_0\sin\phi_0\,d
         -2r_0\delta_{m}\cos\phi_0\,d\,},
\label{eq:r0}
\end{align}
where
\[
\delta_n = n - \frac{M_x+1}{2},\;
\delta_m = m - \frac{M_z+1}{2},\;
\mathbf e_{m,n} = \begin{bmatrix} \delta_n d \\[4pt] 0 \\[4pt] \delta_m d \end{bmatrix},\;
 m=1,\dots,M_z,\; n=1,\dots,M_x.
\]
We use the exact per-element phase but approximate the amplitude by the center distance to get a point to point (P2P) line-of sight (LoS) channel model from the $(m,n)$-th antenna element to the user:
\[
g_{m,n}^{(0)}
\approx \frac{4\pi}{\lambda r_0}\,
\exp\!\Big(-j\frac{2\pi}{\lambda}\,r_{0}^{\,m,n}\Big).
\]
Suppose $\mathbf s_\ell$ is the coordinate of the \(\ell\)-th scatterer and the distance from this scatter to the $(m,n)$-th antenna element is $r_{\ell}^{\,m,n}$, then a similar calculation can be done as in \Cref{eq:r0}.

Define
\[
r_{\ell,0}=\|\mathbf s_\ell\|~~(\text{array-center-to-scatterer distance}) ~~\text{and}~~
r_{\ell,1}=\|\mathbf p-\mathbf s_\ell\|~~(\text{scatterer-to-user distance}).
\]
Then for the non-line-of-sight (NLoS) path, the $r_{\ell,1}$ path will introduce extra path loss and phase delay compared to the LoS path:
\[
g_{m,n}^{(\ell)}
\approx \frac{4\pi}{\lambda r_{\ell,0}\,r_{\ell,1}}
\exp\!\Big(-j\frac{2\pi}{\lambda}\,(r_{\ell}^{\,m,n}+r_{\ell,1})\Big).
\]

Collecting paths and optionally absorbing the approximate amplitude into path coefficients yields the convenient representation
\(
\mathbf h \;=\; \sum_{\ell=0}^{L-1} \beta_\ell\,\tilde{\mathbf g}^{(\ell)},\text{where} \;
\tilde g_{m,n}^{(\ell)}=\exp\!\Big(-j\frac{2\pi}{\lambda}\,r_{\ell}^{\,m,n}\Big),
\)
and the scalar path gain \(\beta_\ell\) is chosen as
\(
\beta_0=\frac{4\pi}{\lambda r_0},\;
\beta_\ell=\frac{4\pi }{\lambda r_{\ell,0}\,r_{\ell,1}}\exp\!\Big(-j\frac{2\pi}{\lambda}\,r_{\ell,1}\Big) (\ell\ge1).
\)
\subsection{Beam Training based on Compressive Sensing}\label{subsec: beam training}
We introduce compressive-sensing-based beam training and the time division for beam training and data transmission in one frame in this section. Suppose the channel can be sparsely represented by a DFT basis. Then, for a DFT codebook matrix $\mathbf{F}$, we have
\begin{equation}
    \mathbf{h}=\mathbf{F}\alpha,
    \label{eq:channel}
\end{equation}
where the implementation of the DFT matrix can be found in \Cref{app: DFT}. In our implementation, we take the DFT grid sizes equal to the UPA dimensions (i.e., $\mathbf{F}\in \mathbb C^{M\times M}$) and the sensing matrix $\Phi$ is realized as a Gaussian mixing applied on the DFT codebook matrix. Concretely, let
\[
\Phi \;=\; \mathbf F\,G^{H}\in\mathbb C^{M\times m},~~ \text{where}~~
G\in\mathbb C^{\,m\times M}~~ \text{and}~~
G_{ij}\sim\mathcal{CN}(0,1/m).
\]
For a pilot vector \(\mathbf x\in\mathbb C^{m}\) the scalar observation is
written as (using \Cref{eq:channel})
\[
y \;=\; \mathbf h^{H}\Phi\mathbf x + w
    \;=\; (\mathbf F\alpha)^{H}\Phi\mathbf x + w
    \;=\; \alpha^{H}\mathbf A\mathbf x + w,
\]
where $w\sim\mathcal{CN}(0,\sigma)$ is additive gaussian noise with power $\sigma^2$. Specializing to canonical per-pilot transmissions \(\mathbf x=\mathbf e_i\) yields
\(y_i=\alpha^{H}\mathbf A\mathbf e_i + w_i=\alpha^{H}\boldsymbol a_i+w_i\) with
\(\boldsymbol a_i\) the \(i\)-th column of \(\mathbf A\). Stacking the \(m\)
measurements as a row vector gives the compact row-form $\mathbf y \;=\; \alpha^{H}\mathbf A + \mathbf w,
~~ \mathbf y\in\mathbb C^{1\times m}.$

We use a revised Target-sparsity Subspace Pursuit (TSP) for compressive sensing \cite{DaiMilenkovic2009_SubspacePursuit} which takes $\mathbf y$, $\mathbf A$ and sparsity level $k$ as input and outputs the recovered coefficients $\hat{\alpha}$.
The detailed algorithm can be found in \Cref{app: beam training}. According to compressive sensing theory, a sufficient condition for exact recovery is that the number of pilot measurements $m$ scales with the sparsity level $k$ \cite{compressive_sensing_RIP}. Because the sparsity of a near-field channel in a DFT dictionary is not fixed and can vary over time and geometry, we let a DRL controller adaptively choose both the sensing dimension (pilot budget $m$) and the working sparsity level $k$ online, rather than fixing them a priori.

From the recovered coefficients \(\hat\alpha\) we form the channel estimate $\hat{\mathbf h} \;=\; \mathbf F\,\hat\alpha \in\mathbb C^{M},$ and use the normalized estimate as a Maximum Ratio Transmission (MRT) precoder
\(
    \hat{\mathbf v} \;=\; \frac{\hat{\mathbf h}}{\|\hat{\mathbf h}\|_2}.
\)
The data-phase receive model is thus
\begin{equation}\label{eq:rec signal}
    y \;=\; \mathbf h^{H}\hat{\mathbf v}\,x + w.
\end{equation}

\subsection{Data Queuing Model}\label{subsec:data_queuing_model}

\begin{figure}[h]
  \centering
 \includegraphics[width=4in]{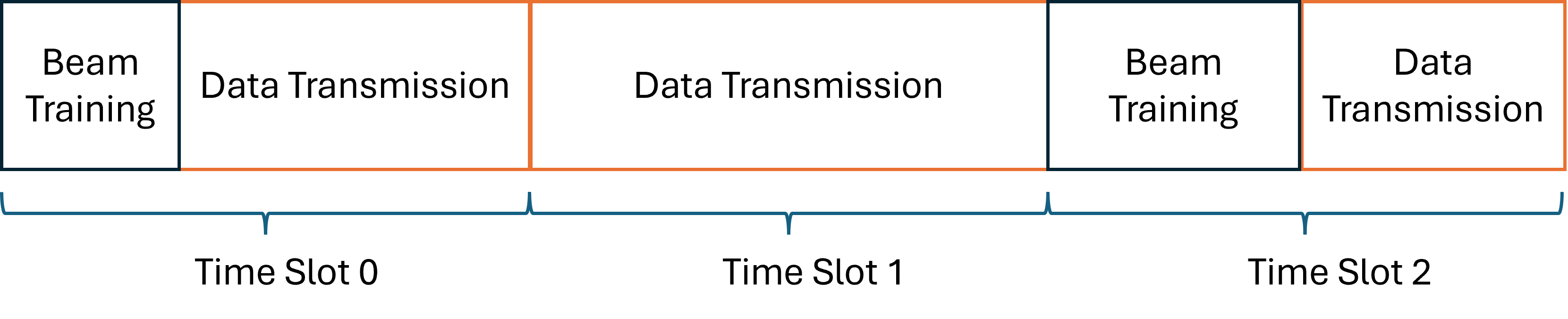}
  \caption{Demonstration of behavior in each time slot.}
  \label{fig: time}
\end{figure}

In this section, we specify the queuing dynamics in each time slot. Suppose the slot duration is \(T_s\). Owing to user mobility, the channel varies across consecutive slots. In each time slot, beam training is performed first (if needed) and then data is transmitted based on the estimated channel. The time that beam training consumes is proportional to the number of pilot measurements in our frame structure, which is illustrated in \Cref{fig: time}. Denote by \(m_t\) the number of pilot measurements used in slot \(t\) and by \(\tau_{\mathrm{ov}}\) the per-pilot overhead (seconds per pilot). The training duration in slot \(t\) is therefore
\begin{equation}\label{eq:Ttrain}
T_{\mathrm{train},t} \;=\; m_t\,\tau_{\mathrm{ov}}.
\end{equation}
The remaining time in the slot is available for data transmission:
\begin{equation}\label{eq:Tdata}
T_{\mathrm{data},t} \;=\; T_s - T_{\mathrm{train},t}.
\end{equation}

According to \Cref{eq:rec signal}, let $\mathrm{SNR}_t=p_t|\mathbf{h}^H\mathbf{v}|^2/\sigma^2$ denote the instantaneous received signal-to-noise ratio (SNR), where $p_t$ is the power for data transmission at the gNB. Then, \(R_t=\log_2(1+\mathrm{SNR}_t)\) denotes the achievable rate in the data phase. With system bandwidth \(W\) (Hz) the number of bits that can be delivered in the data phase of slot \(t\) is
\begin{equation}\label{eq:psi}
\psi_t \;=\; T_{\mathrm{data},t}\,WR_t.
\end{equation}

Let \(q_t\) be the queue length (bits) at the beginning of slot \(t\). Let \(\ell_t\) be the new arrivals (bits) that arrive within slot \(t\), which follows a Poisson Distribution. The buffer has finite capacity \(Q_{\max}\); any excess arrivals that would push the buffer beyond \(Q_{\max}\) are dropped. The queue update is thus written as:
\begin{equation}\label{eq:queue_update}
q_{t+1} \;=\; \min\Big\{Q_{\max},\ \max\{q_t-\psi_t,0\} + \ell_t\Big\},
\end{equation}
and the overflows (dropped bits) in slot \(t\) can be expressed as
\begin{equation}\label{eq:overflow_def}
d_t \;=\; \max\big\{\,q_t - \psi_t + \ell_t - Q_{\max}\,\big\}.
\end{equation}

\section{Problem Formulation}\label{sec:ppo}
In this section, we introduce the proposed cross-layer decision model. 

We model the decision problem as a partially observable Markov decision process (POMDP) \cite{astrom1965pomdp} and solve it with Proximal Policy Optimization (PPO) \cite{schulman2017ppo} as given in \Cref{app: ppo}. To jointly decide the beam training and data transmission procedure, the model needs to output the following actions
\(
a_t \;=\; \big(b_t,\ m_t,\ K_t,\ p_t\big),\;
\)
where \(b_t\in\{0,1\}\) indicates whether CS training is performed in slot $t$, \(m_t\) is the pilot budget (used only if \(b_t=1\)), \(K_t\) is the TSP target sparsity, and \(p_t\) is the data-phase transmit power.

Since we jointly consider the PHY and MAC design, the agent cannot observe the true instantaneous channel state before making a decision. This is in contrast to prior work in which the channel state is assumed to be known \cite{sharma_2020_DRLscheduling5G,Anjali2023scheduling5G}. In this paper, the agent observes a compact tuple that summarizes queueing and recent history. Denote the agent's observation at the start of slot \(t\) by
\(
s_t \;=\; \big(q_t,\ \tau_t,\ I_t\big),
\)
where \(q_t\) is the queue length and \(\tau_t\) is the age (slots since last training). To avoid a degenerate policy that never learns, we enforce $b_t=1$ at $t=0$ or whenever $\tau_t>\tau_{\mathrm{train}}$. A reward-design alternative could eliminate this heuristic, but we leave that as future work.
 \(I_t\) is a short history window of recent $T_\mathrm{age}$ measurement and training tuples:
\begin{equation}
    I_t = \big((b_{t-T_\mathrm{age}},m_{t-T_\mathrm{age}},K_{t-T_\mathrm{age}},R^\prime_{t-T_\mathrm{age}}),\cdots,(b_{t-1},m_{t-1},K_{t-1},R^\prime_{t-1})\big),
    \label{eq:history}
\end{equation}
where $R^\prime=\log_2(1+|\mathbf{h}^H\mathbf{v}|^2/\sigma^2)$ is the original rate without influence of the transmission power.
We obtain transition samples by observing the observation \(s_t\), executing the action \(a_t\) chosen by PPO in the simulator (or system) and observing the resulting next observation \(s_{t+1}\) and reward \(r_t\). Concretely, starting from \(s_t\) and applying \(a_t\) we first compute the training and data transmission durations via \Cref{eq:Ttrain} and \Cref{eq:Tdata}, respectively. If \(b_t=1\) the environment returns measurements \(\mathbf y_t\) and the TSP recovery \(\hat\alpha_t\), from which the channel estimate \(\hat{\mathbf h}_t\) and the beamformer are constructed. The data-phase SNR and delivered service are then computed via \Cref{eq:psi}, and the next queue state is determined via \Cref{eq:queue_update}. 

We design the immediate reward to reflect energy, delay, and overflow costs. The beam training process uses maximum power (normalized to 1) for beam sweeping. Define the per-slot energy as
\begin{equation}\label{eq:energy_perslot}
E_t \;=\; E_{\mathrm{train}}(m_t) \;+\; p_t\,T_{\mathrm{data},t}
~~ \text{and} ~~
E_{\mathrm{train}}(m_t)=m_t\tau_{\mathrm{ov}},
\end{equation}
where $m_t$ is the pilot budget, $\tau_{\mathrm{ov}}$ is the time consumption of one single beam for beam sweeping, $p_t\in[0,1]$ is the normalized transmit power, and $T_{\mathrm{data},t}$ is given by \Cref{eq:Tdata}. The per-slot reward is then
\begin{equation}\label{eq:reward}
r_t \;=\; -( E_t \;+\; \lambda_Q\,q_{t+1} \;+\; \lambda_{\mathrm{drop}}\,d_t ),
\end{equation}
with $\lambda_Q>0$ weighting the delay penalty via the next-queue length $q_{t+1}$ from \Cref{eq:queue_update}, and $\lambda_{\mathrm{drop}}\ge 0$ weighting the overflow penalty through the dropped bits $d_t$ in \Cref{eq:overflow_def}. Thus, \Cref{eq:reward} is equivalently a cost $c_t=E_t+\lambda_Q q_{t+1}+\lambda_{\mathrm{drop}} d_t$ with $r_t=-c_t$, where the energy term captures training and data-phase expenditure and the queueing terms capture latency and reliability consequences of the current decision $(b_t,m_t,K_t,p_t)$.

\section{Experiments}\label{sec:experiments}
In this session, we provide the simulation results and discussion. To evaluate the proposed model, we built a custom Gymnasium \cite{towers2024gymnasium} environment based on the system model presented in \Cref{sec:system model}. Our learning agent is based on StableBaseline3 PPO \cite{stable-baselines3}.
 We discretize continuous components (e.g., \(p_t\)) and use a shared MLP trunk with separate policy/value heads. The central frequency is $f_c=$ 30 GHz and antenna size is $M_x=128,M_z=8$. The detailed parameter selection, component discretization, and experiment setting can be found in \Cref{app: parameters}. Let an episode contain $T_{\mathrm{tol}}$ slots. We report the following metrics, and first form episode-level quantities, then report across-episode mean $\pm$ standard deviation.

\textbf{Achievable rate.}
The per-episode mean rate is
$ {\tilde R} = (1/T_\mathrm{tol})\sum_{t=1}^{T_\mathrm{tol}}\, \psi_t / T_s$ (bps).

\textbf{Beamforming-gain ratio.}
Per-slot
$\rho_t = |\mathbf{v}_t^{H}\mathbf{h}_t|^2 / \|\mathbf{h}_t\|^2 \in [0,1]$, with time average
$\rho = (1/T_\mathrm{tol})\sum_{t=1}^{T_\mathrm{tol}}\rho_t$.

\textbf{Overflow rate.}
With the queue update in \Cref{eq:queue_update}, and per-slot overflow $d_t$ defined in \Cref{eq:overflow_def}, we report the bits-based overflow fraction over $T_\mathrm{tol}$ slots as
\(
\mathrm{Ov}(\%) = 100 \times \bigl(\sum_{t=1}^{T_\mathrm{tol}} d_t\bigr)\, / \,\bigl(\sum_{t=1}^{T_\mathrm{tol}} \ell_t\bigr),
\)
where $\ell_t$ is the number of arrived bits in slot $t$.

\textbf{Energy consumption.}
The mean energy is
$ E = (1/T_\mathrm{tol})\sum_{t=1}^{T_\mathrm{tol}} E_t$.

\textbf{Train time fraction.}
Using the per-slot training duration \(T_{\mathrm{train},t}\) from \Cref{eq:Ttrain} and the slot length \(T_s\) from \Cref{eq:Tdata}, we report
\(
\mathrm{TTF}(\%) \;=\; 100 \times (1/T_\mathrm{tol})\sum_{t=1}^{T_\mathrm{tol}}\bigl(T_{\mathrm{train},t}/T_s\bigr),
\)





\subsection{Comparison with Baselines}\label{subsec:baseline}
We compare our method against three baselines.

\textbf{5G NR (DFT codebook).}
Following 5G NR beam management, we perform beam sweeping over a DFT codebook and select the beam with the highest instantaneous SNR for data transmission. Beam training is executed periodically every $\tau_{\mathrm{train}}$ slots. The data transmission power is fixed at the same maximum power $P_{\max}$ used by our method.

\textbf{Near-field–improved 5G NR.}
This baseline uses the same training schedule and power setting as above but replaces the DFT codebook with the near-field codebook from \cite{wuMultipleAccessNearField2023}, which extends \cite{Cui_and_Dai_channel_model} to UPA.

\textbf{Full CSI (oracle upper bound).}
We assume perfect channel knowledge and use the matched filter/maximum-ratio beam
\(
\mathbf{v}^{\star}_t = \mathbf{h}_t/\lVert\mathbf{h}_t\rVert.
\)
No beam training is needed (zero time/energy overhead) and data power is set to $P_{\max}$. This serves as an upper bound on performance.


\begin{table}[h]
\centering
\setlength{\tabcolsep}{10pt}
\renewcommand{\arraystretch}{0.95}
\caption{Baseline comparison.}
\label{tab:baseline-comparison}
\begin{tabular}{@{}lccccc@{}}
\toprule
Method      & $\bar{\tilde{R}} $ (Mbps) & $\bar\rho$ & $\overline{\mathrm{Ov}}$ (\%) & $\bar E$ & $\overline{\mathrm{TTF}}$ (\%) \\
\midrule
DFT        & \(26.9\pm4.1\)  & \(0.317\pm0.104\) & \(6.7\pm4.7\) & \(0.001\pm0.000\) & \(5.0\pm0.0\) \\
Near-field  & \(31.0\pm6.3\)  & \(0.410\pm0.101\) & \(9.3\pm8.9\) & \(0.001\pm0.000\) & \(9.3\pm0.0\) \\
Full CSI    & \(57.1\pm7.3\)  & \(1.000\pm0.000\) & \(0.4\pm2.2\) & \(0.001\pm0.000\) & \(0.0\pm0.0\) \\
Proposed    & \(49.9\pm34.0\) & \(0.853\pm0.052\) & \(1.5\pm6.1\) & \(0.001\pm0.000\) & \(3.2\pm0.0\) \\
\bottomrule
\end{tabular}
\end{table}

Relative to the two 5G-NR style baselines, the proposed method increases throughput to \(49.9\) Mbps, which corresponds to a \(+85.5\%\) improvement over DFT sweeping at \(26.9\) Mbps and \(+61.0\%\) over the near-field codebook at \(31.0\) Mbps. 
The beamforming-gain ratio is \(\rho=0.853\), which is \(2.69\times\) the DFT value \(0.317\) and \(2.08\times\) the near-field value \(0.410\).
Queueing performance improves accordingly: the overflow rate is reduced to \(1.5\%\), i.e., a decrease of \(5.2\) percentage points relative to DFT (\(6.7\%\)) and \(7.8\) percentage points relative to near-field (\(9.3\%\)).
Training overhead is also lower at \(3.2\%\), compared with \(5.0\%\) for DFT and \(9.3\%\) for near-field.
A gap to the full-CSI upper bound remains: the proposed rate achieves \(49.9/57.1\approx87.4\%\) of the oracle throughput.
Energy consumption is identical across methods in this setup because all policies transmit at maximum power (as in the 5G-NR baselines). Exploring the energy–delay trade-off by adjusting the cost weights in \Cref{eq:reward} is left for future work.

\subsection{Ablation Experiments}\label{subsec:ablation}
We conduct two ablations to quantify the contribution of temporal history in the observation and periodic beam training. In \textbf{No history}, we remove the history by setting $T_{\mathrm{age}}=0$. In \textbf{No train}, we eliminate periodic training and train only at $t=0$ (i.e., no $\tau_{\mathrm{train}}$). The full model (\textbf{w/o ablation}) uses both components.

\begin{table}[h]
\centering
\setlength{\tabcolsep}{9pt}
\renewcommand{\arraystretch}{0.95}
\caption{Ablation comparison.}
\label{tab:ablation-comparison}
\begin{tabular}{@{}lccccc@{}}
\toprule
Method        & $\bar{\tilde{R}}$ (Mbps) & $\bar\rho$ & $\overline{\mathrm{Ov}}$ (\%) & $\bar E$ & $\overline{\mathrm{TTF}}$ (\%) \\
\midrule
No history    & \(36.8\pm5.3\)  & \(0.566\pm0.065\) & \(5.3\pm5.4\)  & \(0.001\pm0.000\) & \(3.6\pm0.0\) \\
No train      & \(35.2\pm23.8\) & \(0.374\pm0.083\) & \(6.4\pm12.4\) & \(0.001\pm0.000\) & \(0.0\pm0.0\) \\
w/o ablation  & \(49.9\pm34.0\) & \(0.853\pm0.052\) & \(1.5\pm6.1\)  & \(0.001\pm0.000\) & \(3.2\pm0.0\) \\
\bottomrule
\end{tabular}
\end{table}

Energy consumption is identical across methods in this setup.
The full model (w/o ablation) achieves \(49.9\) Mbps with \(\rho=0.853\) and an overflow rate of \(1.5\%\) at \(3.2\%\) training time. Removing history reduces the rate to \(36.8\) Mbps, a \(26.3\%\) drop relative to the full model, and lowers beamforming-gain ratio to \(\rho=0.566\) (a \(33.6\%\) decrease). The overflow rate increases by \(3.8\) percentage points to \(5.3\%\); training time is slightly higher at \(3.6\%\). Eliminating periodic training is more detrimental: the rate falls to \(35.2\) Mbps (\(29.5\%\) below full), beamforming-gain ratio drops to \(\rho=0.374\) (\(56.1\%\) decrease), and the overflow rate rises by \(4.9\) percentage points to \(6.4\%\) while using \(0\%\) training time. These results indicate that both temporal context and periodic training contribute to performance, with training providing the larger share of the gain in both throughput and alignment.

\section{Conclusion}\label{sec:con}
We presented a cross-layer design for near-field mmWave tailored to delay-sensitive traffic under explicit energy constraints. By jointly optimizing beam alignment, power, and queue-aware scheduling, our method balances training overhead and data transmission time, yielding higher achievable rate and lower overflow than DFT and near-field codebook baselines while approaching full-CSI performance. Future work includes extending to multi-user settings, formulating simpler greedy methods to reduce the model training and evaluation time, exploiting ML optimization algorithms with higher sample efficiency and lower training convergence time \cite{SharmaMC:18a,felizardo2025reinforcementlearningmethodenvironments}, and integration of application-centric objectives for further impact on next generation wireless IoT and XR application systems.

\section*{Acknowledgments}
This material is based upon work supported in part by the National Science Foundation (NSF) under Grant Nos. CNS-2106150, CNS-2032033, CNS-2346528, and ECCS-2512911.

{
\bibliographystyle{nips} 
\bibliography{ref}          
}


\appendix

\section{DFT codebook implementation}\label{app: DFT}
In our implementation, we set the DFT grid sizes equal to the UPA dimensions, i.e., \(P=M_x\) and \(Q=M_z\). Let \(M=M_xM_z\) and stack the UPA elements into the column vector \(\mathbf h\in\mathbb C^{M}\) using column-major order (index \(m\) varies fastest inside each \(n\)). Recall \(\delta_n\) and \(\delta_m\) as defined above. Define the 1-D spatial frequency grids
\[
u_p=\frac{2p}{M_x}-1,\quad p=0,\dots,M_x-1,
\qquad
w_q=\frac{2q}{M_z}-1,\quad q=0,\dots,M_z-1.
\]
For each grid pair \((u_p,w_q)\) the corresponding normalized steering column \(\mathbf f_{p,q}\in\mathbb C^{M}\) has entries
\[
\mathbf f_{p,q}[m,n]
=\frac{1}{\sqrt{M}}\exp\!\big(-j\pi(\delta_n u_p + \delta_m w_q)\big),
\qquad m=1,\dots,M_z,\; n=1,\dots,M_x,
\]
where \(\mathbf f_{p,q}[m,n]\) denotes the entry at element \((m,n)\) in the same stacking order used for \(\mathbf h\). The 2-D DFT dictionary (codebook) is the \(M\times M\) matrix formed by concatenating these columns,
\[
\mathbf F = \big[\,\mathbf f_{0,0}\;\mathbf f_{1,0}\;\dots\;\mathbf f_{M_x-1,M_z-1}\,\big]\in\mathbb C^{M\times M}.
\]
\section{Main system parameters}\label{app: parameters}
We evaluate all methods(\Cref{sec:system model}), each run for \(E{=}40\) episodes of \(T{=}5000\) slots under distinct random seeds (\(\texttt{seed}_0+\text{episode index}\)). Unless otherwise stated, policies are evaluated in deterministic (greedy) mode (\texttt{deterministic=True} in PPO), and all PHY/MAC/system parameters are kept identical across methods. Baselines use the settings in \Cref{subsec:baseline} (DFT or near-field codebook sweeping every \(\tau_{\mathrm{train}}\) slots with \(p_t{=}P_{\max}\); full-CSI uses matched filtering with no training overhead).
\begin{table}[h]
\centering
\caption{Main system parameters.}
\label{tab:parameter}
\begin{tabularx}{\linewidth}{l X X}
\toprule
\thead{Parameter} & \thead{Meaning} & \thead{Value} \\
\midrule
$f_c$ & Carrier (central) frequency & $30\,\mathrm{GHz}$ \\
$\lambda$ & Wavelength ($\lambda = c/f_c$) & $10\,\mathrm{mm}$ \\
$d$ & Antenna spacing & $\lambda/2 = 5\,\mathrm{mm}$ \\
$M_x\times M_z$ & UPA size (horizontal $\times$ vertical) & $128\times 8$ \;($M=1024$ elements) \\
$W$ & System bandwidth & $20\,\mathrm{MHz}$ \\
$T_s$ & Slot duration & $1\,\mathrm{ms}$ \\
$\tau_{\mathrm{train}}$ & Beam-training period & $10$ slots (default) \\
$\tau_{\text{ov}}$ & Time for one overhead in beam training & $1/2048\times 10^{-4}\,\mathrm{ms}$ \\
$L_{\text{path}}$ & Number of channel paths & $3$ \\
$\sigma^2$ & Noise variance & $5.2\times 10^{-10}$ \\
$T_{\text{age}}$ & History window length in observation & $64$ slots \\
$Q_{\max}$ & Queue capacity & $80{,}000$ bits \\
$\lambda_{\text{arr}}$ & Packet arrival intensity & $2000$ pkts/s \\
$b_{\text{pkt}}$ & Bits per packet & $6000$ bits \\
$\mathbb{E}[\ell_t]$ & Mean arrival per slot ($\lambda_{\text{arr}} b_{\text{pkt}} T_s$) & $12{,}000$ bits/slot \\
$T$ & Maximum iterations for TSP & 1 \\
$P_{\text{level}}$ & Power discretization levels & $10$ (i.e., $p\in\{0.1,\dots,1.0\}, P_\mathrm{max}=1$) \\
$m_{\text{level}}$ & Pilot/overhead budget levels 
& $10$ (ratios; $m_i=\big\lceil\tfrac{i+1}{m_{\text{level}}}M \big\rceil, i=0,\dots,9$, with $M=M_xM_z$) \\

$K_{\text{level}}$ & Sparsity levels 
& $10$ (ratios; $K_i=\max\{1,\big\lceil \tfrac{i+1}{2K_{\text{level}}}m_t \big\rceil\},i=0,\dots,9$) \\

\bottomrule
\end{tabularx}
\end{table}

\section{Compressive sensing based beam training}\label{app: beam training}
For algorithmic recovery we equivalently work with the column-form by taking
Hermitian transpose:
\[
\tilde{\mathbf y} \;=\; \mathbf y^{H} \in\mathbb C^{m\times 1},\qquad
\tilde{\mathbf A} \;=\; \mathbf A^{H} \in\mathbb C^{m\times M},\qquad
\tilde{\mathbf w} \;=\; \mathbf w^{H} \in\mathbb C^{m\times 1}
\]
which yields the standard CS model
\[
\tilde{\mathbf y} \;=\; \tilde{\mathbf A}\,\alpha + \tilde{\mathbf w}.
\]
The detailed algorithm is in \Cref{alg:SPK_left}.
\begin{algorithm}[h]
\caption{TSP }
\label{alg:SPK_left}
\begin{algorithmic}[1]
\REQUIRE $\mathbf A\in\mathbb C^{M\times m}$, row-measurements $\mathbf y\in\mathbb C^{1\times m}$, target sparsity $k$, max iterations $T$
\STATE $\tilde{\mathbf y}\leftarrow \mathbf y^{H}$,\quad $\tilde{\mathbf A}\leftarrow \mathbf A^{H}$
\STATE $\mathbf z \leftarrow \tilde{\mathbf A}^{H}\tilde{\mathbf y}$ 
\STATE $\widehat{\mathcal S}\leftarrow \operatorname{TopK}(|\mathbf z|,k)$  (TopK: indices of the $k$ largest $|\mathbf z_i|$)
\FOR{$t=1$ \textbf{to} $T$}
  \STATE $\hat\alpha_{\widehat{\mathcal S}} \leftarrow \tilde{\mathbf A}_{\widehat{\mathcal S}}^\dagger \tilde{\mathbf y}$  (LS on current support (pseudoinverse))
  \STATE $\mathbf r \leftarrow \tilde{\mathbf y} - \tilde{\mathbf A}_{\widehat{\mathcal S}}\hat\alpha_{\widehat{\mathcal S}}$
  \STATE $\mathbf z \leftarrow \tilde{\mathbf A}^{H}\mathbf r$ 
  \STATE $\widehat{\mathcal S}_{\text{new}} \leftarrow \widehat{\mathcal S}\ \cup\ \operatorname{TopK}(|\mathbf z|,k)$
  \IF{$\widehat{\mathcal S}_{\text{new}} = \widehat{\mathcal S}$}
    \STATE \textbf{break}  (Converged: support unchanged)
  \ENDIF
  \STATE $\widehat{\mathcal S}\leftarrow \widehat{\mathcal S}_{\text{new}}$
\ENDFOR
\STATE $\hat\alpha \leftarrow \mathbf 0\in\mathbb C^{M}$; \quad $\hat\alpha_{\widehat{\mathcal S}} \leftarrow \tilde{\mathbf A}_{\widehat{\mathcal S}}^\dagger \tilde{\mathbf y}$
\RETURN $\hat\alpha,\ \widehat{\mathcal S}$
\end{algorithmic}
\end{algorithm}

\section{PPO}\label{app: ppo}
We denote policy and value parameters by \(\boldsymbol{\omega}\) and \(\boldsymbol{\nu}\), respectively.

We estimate advantages with generalized advantage estimation (GAE) using a critic \(V_{\boldsymbol{\nu}}\):
\begin{equation}\label{eq:delta}
\delta_t \;=\; r_t + \gamma V_{\boldsymbol{\nu}}(s_{t+1}) - V_{\boldsymbol{\nu}}(s_t),\qquad
\hat A_t \;=\; \sum_{\ell=0}^{L-1}(\gamma\,\lambda_{\mathrm{GAE}})^{\ell}\,\delta_{t+\ell}.
\end{equation}
The multi-discrete action \(a_t=(b_t,m_t,K_t,p_t)\) is modeled by a factorized categorical policy
\begin{equation}\label{eq:policy_discrete}
\pi_{\boldsymbol{\omega}}(a_t\mid s_t) \;=\; \prod_{j=1}^{4}\pi_{\boldsymbol{\omega}}^{(j)}\!\big(a_t^{(j)}\mid s_t\big).
\end{equation}
PPO maximizes the clipped surrogate with ratio \(\rho_t(\boldsymbol{\omega})=\tfrac{\pi_{\boldsymbol{\omega}}(a_t\mid s_t)}{\pi_{\boldsymbol{\omega}_{\mathrm{old}}}(a_t\mid s_t)}\):
\begin{equation}\label{eq:ppo_clip}
\mathcal{L}^{\mathrm{CLIP}}_t(\boldsymbol{\omega})=\min\!\Big(\rho_t(\boldsymbol{\omega})\,\hat A_t,\ \mathrm{clip}\big(\rho_t(\boldsymbol{\omega}),1-\epsilon,1+\epsilon\big)\,\hat A_t\Big),
\end{equation}
and we train an actor–critic with
\begin{equation}\label{eq:ppo_loss}
L(\boldsymbol{\omega},\boldsymbol{\nu}) \;=\; -\mathbb{E}_t[\mathcal{L}^{\mathrm{CLIP}}_t] 
\;+\; c_1\,\mathbb{E}_t\!\big[\big(V_{\boldsymbol{\nu}}(s_t)-r^\prime_t\big)^2\big]
\;-\; c_2\,\mathbb{E}_t\!\big[\mathcal{H}\big(\pi_{\boldsymbol{\omega}}(\cdot\mid s_t)\big)\big],
\end{equation}
where \(c_1,c_2>0\) weight value regression and entropy regularization. The target \(r'_t\) is the truncated, bootstrapped return
Rollouts are used to compute \(\hat A_t\) in \Cref{eq:delta} and to optimize \Cref{eq:ppo_loss} via minibatch SGD.

\paragraph{Training schedule (total rounds).}
Unless otherwise noted, we use $\gamma=0.99$ and $\lambda_{\mathrm{GAE}}=0.95$.
Training proceeds for $N_{\mathrm{upd}}$ PPO updates, each collecting $n_{\mathrm{steps}}$ transitions per environment over $N_{\mathrm{env}}$ parallel environments.
The total number of environment steps is
\[
T_{\mathrm{tot}} \;=\; N_{\mathrm{upd}}\; n_{\mathrm{steps}}\; N_{\mathrm{env}}.
\]
In our reported runs we used $T_{\mathrm{tot}}=15.16\text{M}$ environment steps.
Each update performs $K_{\mathrm{epoch}}$ epochs of minibatch SGD with minibatch size $M_{\mathrm{mb}}$.
The remaining hyperparameters are standard PPO: clipping parameter $\epsilon$, learning rate $\eta$ with Adam, value loss weight $c_1$, and entropy weight $c_2$.

\paragraph{Network architecture.}
Observations are fed to a shared multilayer perceptron (MLP) with two hidden layers of widths $(128,64)$ and elementwise nonlinearity (ReLU or Tanh).
From the shared trunk, we branch into:
(i) a policy head that outputs concatenated logits for the four categorical factors
\[
\underbrace{2}_{b_t}\;+\;\underbrace{m_{\mathrm{level}}}_{m_t}\;+\;\underbrace{K_{\mathrm{level}}}_{K_t}\;+\;\underbrace{P_{\mathrm{level}}}_{p_t}\;=\;32,
\]
which are then partitioned to form $\pi_{\boldsymbol{\omega}}^{(j)}$ in \Cref{eq:policy_discrete}; and
(ii) a value head that outputs the scalar $V_{\boldsymbol{\nu}}(s_t)$.
The input size (observation size) for the full model is $4T_{\mathrm{age}}+2$ (the $4$ features per slot from the history window plus $q_t$ and $\tau_t$); with $T_{\mathrm{age}}=64$ this equals $258$.
For the no-history ablation we set $T_{\mathrm{age}}=0$, giving an input size of $4\cdot 0+2=2$.
Both actor and critic use the shared $(128,64)$ trunk with separate linear output layers, advantage normalization, and gradient clipping.
\end{document}